\documentclass[10pt]{iopart}
\usepackage{times}
\usepackage{iopams}
\usepackage{mathptmx}
\usepackage[dvips]{graphicx}
\newcommand{\ped}[1]{\ensuremath{_{\rm #1}}}

\begin{document}

\title[Point-Contact Spectroscopy in MgB$_2$]
{Point-Contact Spectroscopy in MgB$_2$: from Fundamental
Physics to Thin-Film Characterization}

\author{R.S. Gonnelli \dag\footnote[6]{To whom correspondence should be addressed
(renato.gonnelli@polito.it)}, D. Daghero~\dag, A. Calzolari~\dag,
G.A. Ummarino~\dag, V. Dellarocca~\dag, V.A. Stepanov~\ddag, S.M.
Kazakov~\S, J. Karpinski~\S, C. Portesi~$\|$, E. Monticone~$\|$,
V. Ferrando~$\P$ and C. Ferdeghini~$\P$}

\address{\dag INFM -- Dipartimento di Fisica, Politecnico di Torino,
10129 Torino, Italy}

\address{\ddag P.N. Lebedev Physical Institute, Russian Academy of
Sciences, 119991 Moscow, Russia}

\address{\S Solid State Physics Laboratory, ETH, CH-8093
Z\"{u}rich, Switzerland}

\address{$\|$ Istituto Elettrotecnico Nazionale ``Galileo Ferraris'', 10135 Torino, Italy}

\address{$\P$ INFM-LAMIA and Dipart. di Fisica, Universit\`{a} di Genova, 16146 Genova}

\begin{abstract}
In this paper we highlight the advantages of using point-contact
spectroscopy (PCS) in multigap superconductors like MgB$_2$, both
as a fundamental research tool and as a non-destructive diagnostic
technique for the optimization of thin-film characteristics. We
first present some results of crucial fundamental interest
obtained by directional PCS in MgB$_2$ single crystals, for
example the temperature dependence of the gaps and of the critical
fields and the effect of a magnetic field on the gap amplitudes.
Then, we show how PCS can provide useful information about the
surface properties of MgB$_2$ thin films (e.g. $T\ped{c}$, gap
amplitude(s), clean or dirty-limit conditions) in view of their
optimization for the fabrication of tunnel and Josephson junctions
for applications in superconducting electronics.
\end{abstract}
\section{Introduction}
From the point of view of fundamental physics, magnesium diboride
(MgB$_2$, discovered to be superconducting below 39~K in 2001
\cite{Akimitsu}) is particularly interesting because it is the
clearest example of two-band superconductor ever
studied~\cite{twoband,Brinkman,Choi}. However, this simple
compound is also interesting because of its possible applications,
that could take advantage of its rather low production cost and of
its rather high $T\ped{c}$. At present, one of the most promising
fields of application is superconducting electronics, with the
perspective of MgB$_2$-based devices with good performances
operating at a temperature ($\sim$ 10-15 K) accessible to
one-stage cryocoolers. After the early observation of the
Josephson effect in MgB$_2$ break junctions \cite{PRLJosephson},
the efforts in this directions have led to the fabrication of
various kinds of Josephson and tunnel junctions~\cite{Junctions}
as well as prototypal SQUIDs~\cite{SQUID}.

Point-contact spectroscopy (PCS) has proved particularly useful in
the fundamental research on MgB$_2$, since it allows measuring in
a direct way both the gaps of this compound \cite{szabo} with
great accuracy \cite{nostroPRL}. Actually, being a
surface-sensitive probe, PCS is also a unique experimental
technique to study the surface superconducting properties of the
material. Thus, we also used it as a diagnostic tool within a
process of optimization of MgB$_2$ thin films, in view of the
fabrication of Josephson junctions of different kinds and,
finally, of simple superconducting electronic devices (e.g.
SQUIDs).

In this paper we will first present some results of relevant
fundamental interest we obtained by directional PCS in MgB$_2$
single crystals in the presence of magnetic fields up to 9~T,
among which the temperature dependence of the critical field for
$\mathbf{B}\parallel c$ and $\mathbf{B}\parallel ab$, and the
temperature and magnetic-field dependence of the gap amplitudes.
We will show how these results address some important subjects of
present debate in the fundamental physics of MgB$_2$.
Then, we will report the results of PCS measurements in MgB$_2$
thin films grown by different techniques on various substrates.
These measurements provide information (e.g. about film
orientation) that can be compared to structural and morphological
results, such as  XRD and AFM/STM. We will also show that PCS can
give very useful indications about the conditions of the very
surface (e.g., presence and effectiveness of impurities,
oxidation, degradation) that are of crucial importance for the
fabrication of tunnel and Josephson junctions for electronic
applications, but \emph{cannot} be provided by bulk-sensitive
techniques.

\section{Experimental details}
\subsection{The samples}
The state-of-the-art MgB$_2$ single crystals used in our
experiments were produced by J.~Karpinski's group at the Solid
State Physics Laboratory of ETH, Zurich (Switzerland). The crystal
growth procedure occurs under a pressure of 30-35 kbar and is
described in detail elsewhere (see for example
Ref.~\cite{Karpinski}). For our purposes, we had to use crystals
with regular shape (i.e. flat upper surface and sharp edges) even
if they were rather small ($0.6\times 0.6 \times 0.06 $~mm$^3$ at
most).

The thin films were produced either at the National
Electrotechnical Institute (IEN) ``Galileo Ferraris'' in Turin
(Italy) or at the LAMIA Laboratory of the National Institute for
the Physics of Matter (INFM), in Genoa (Italy).
%
%%%%%%%%% Ferdeghini %%%%%%%%%%%%%%%%%%%%%
Films on $c$-cut sapphire and MgO (111) (both with hexagonal
surface symmetry, similar to that of MgB$_2$) were produced by
using a two-step procedure by the Genoa group \cite{Ferdeghini}.
The first step consists in the PLD deposition of an amorphous,
non-superconducting precursor layer in UHV ($10^{-9}$ mbar) at
room temperature, starting from a stoichiometric MgB$_2$ sintered
target. The second step is an ex-situ annealing process in
magnesium atmosphere, necessary to crystallize the superconducting
phase. The samples are sealed in tantalum crucibles under Ar
pressure with Mg lumps, closed in an evacuated quartz tube and
kept at 850$^\circ$C for half an hour with a following rapid
quenching to room temperature.
%
%%%%%%%%%%%% Monticone %%%%%%%%%%%%%%%%%%%%%%
In addition to sapphire, the group from Turin also used amorphous
silicon nitride (SiN) as substrate, which is particularly
advantageous in view of the fabrication of
bolometers~\cite{Monticone}. The 500~nm-thick, amorphous SiN layer
is grown on silicon wafer by LPCVD. The MgB$_2$ films are then
deposited by simultaneous evaporation of Mg and B at constant
temperature in a pressure of 5$\cdot 10^{-7}$ mbar from a Mo
resistive heater and from a Mo crucible by e-beam heating,
respectively. The resulting precursor is annealed \emph{in situ}
(i.e. in the deposition chamber) in Ar atmosphere, at
500$^\circ$C. The resulting films look golden-brownish and
mirror-like, are always very homogeneous and cover the whole
surface of the substrates (about 5 cm$^2$).

\subsection{Directional PCS in single crystals}
As described elsewhere~\cite{nostroPRL}, the point contacts (of
about $50\times50\,\mu\mathrm{m}^2$) were made by using a small
piece of indium or a drop of Ag conductive paint. This ensured
better mechanical stability on thermal cycling and reproducibility
of the conductance curves with respect to the standard technique
that employs a metallic tip pressed against the sample. The
contacts were placed either on the top surface or on the side of
the crystals, so as to inject the current mainly along the $c$
axis or along the $ab$ planes, respectively~\footnote[7]{Actually,
the current is injected within a cone whose angle $\phi$ depends
on the potential barrier at the interface: $\phi \rightarrow 0$
for tunneling ($Z\rightarrow \infty$) and $\phi=\pi/2$ for
metallic contact with no barrier ($Z=0$). Our contacts are always
in an intermediate case, but the probability for electrons to be
injected along an angle $\varphi$ in the cone is proportional to
$\cos{\varphi}$ \cite{Tanaka} so that it is maximum along the
normal direction anyway.}.

Knowing the direction of current injection is important because of
the anisotropy of MgB$_2$. In fact, PCS gives the $I-V$
characteristics and the conductance curves (d$I$/d$V$ vs. $V$) of
a normal metal/superconductor junction. In the case of MgB$_2$,
the conductance across such a junction contains (separate)
contributions of the $\sigma$ and $\pi$ bands
\cite{Brinkman,nostroPRL}. Since the $\sigma$ bands are almost 2D
(and originate from the overlapping of in-plane $sp^{2}$ boron
orbitals), their contribution to the conductance is maximum for
current flowing along the $ab$ planes. This is thus the most
favorable configuration to observe clearly the $\sigma$-band gap
together with the $\pi$-band one. The use of MgB$_2$ single
crystals allowed us to perform these measurements and to test the
predictions of the two-band model concerning the amplitudes of the
gaps, their temperature- and magnetic field-dependency, and the
relative weight of the two bands for $ab$-plane and $c$-axis
current injection.

\subsection{PCS in thin films}
As in single crystals, point contacts on films were made either by
using a spot of Ag paint or a piece of indium placed on the top
surface~\footnote[6]{Note that the conventional PCS technique
using a metallic tip could be unadvisable in this case, due to the
risk of cracks in the film itself.}. In this case, the weight of
the $\sigma$-band contribution to the total normalized conductance
is an indicator of the film orientation. For perfectly $c$-axis
oriented films an almost pure $c$-axis conductance is expected,
with little or no traces of the $\sigma$-band gap (as an example,
see the lower curve in Figure\ref{Fig1}a). However, even in highly
oriented films the $\sigma$-band contribution could be greater
than expected if the surface roughness is not small enough, since
microcontacts could be established also on the edge of grains or
mesas. Similarly, even in films with very flat surface (such as
those grown by hybrid physical-chemical vapour deposition (HPCVD)
\cite{HPCVD}) conventional PCS can give an excess $\sigma$-band
contribution due to the partial penetration of the tip in the
films. Probably, this explains the recent results by Bugoslavskij
and coworkers \cite{Bugoslavsky}.

In addition to the information about film orientation, PCS also
provides information about possible enhancements or reductions of
the superconducting properties at the surface the clean- or
dirty-limit conditions at the surface (i.e., roughly speaking, the
effectiveness of interband scattering due to impurities) and the
existence of a degraded insulating layer, for example of MgO
(which can arise from the reaction with atmosphere).

\section{Results and discussion}
\subsection{Directional PCS in single crystals}
\begin{figure}[t]
\includegraphics[keepaspectratio, width=\textwidth]{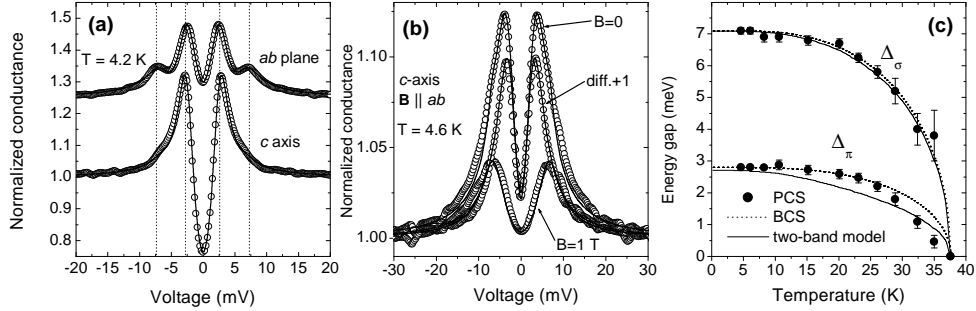}
\vspace{-5mm} \caption{Some experimental results of directional
PCS in MgB$_2$ single crystals. (a) Typical normalized conductance
curves of a $ab$-plane and of a $c$-axis contact. Symbols:
experimental data. Lines: best-fitting curves (two-band BTK
model). (b) Symbols: experimental normalized conductance curves of
a $c$-axis In contact, measured at 4.6 K in zero field (``B=0'')
and in the presence of a field of 1~Tesla (``B=1T''), together
with the difference between these two, vertically shifted by 1 for
best comparison (``diff.+1''). Solid lines: best-fitting curves
obtained by using the functions $\sigma\ped{B=0}$,
$\sigma\ped{B=1T}$ and $\sigma\ped{diff}+1$ defined in the text.
(c) Temperature dependence of the gaps $\Delta\ped{\sigma}$ and
$\Delta\ped{\pi}$ obtained from the fit of the partial $\sigma$-
and $\pi$-band conductances (symbols). Dotted lines: BCS-like
$\Delta(T)$ curves. Solid lines: predictions of the two-band model
in the Eliashberg formulation \cite{Brinkman}.}\label{Fig1}
\end{figure}

Figure~1 reports some results of PCS in single crystals. In panel
(a) symbols represent typical conductance curves of $ab$-plane and
$c$-axis contacts. In the first case (upper curve) four
conductance peaks are present, at $V\simeq\pm 2.8$~mV and $V\simeq
\pm 7.6$~mV. These features are connected to the two gaps,
$\Delta\ped{\pi}$ and $\Delta\ped{\sigma}$. In $c$-axis contacts
(lower curve) at the same voltages we observe clearly defined
maxima and smooth shoulders, respectively. Such a difference is
due to the aforementioned anisotropy of the $\sigma$ bands, whose
contribution to the conductance is minimum for $c$-axis current.
That this contribution is not zero is clearly shown by the fact
that the standard $s$-wave, single-band Blonder-Tinkham-Klapwijk
(BTK) model \cite{BTK} is not able to fit the conductance curves.
On the contrary, the same model generalised to the two-band case,
where the total \emph{normalized} conductance is
$\sigma=w\ped{\pi}\sigma\ped{\pi}+
(1-w\ped{\pi})\sigma\ped{\sigma}$, gives rather good results
(solid lines). The best fit is obtained with values of the gaps
and of the weight $w\ped{\pi}$ \cite{nostroPRL} that agree very
well with the predictions of the two-band model in the Eliashberg
formulation \cite{Brinkman}. However, the accuracy of these values
is rather poor since the fitting function contains 7 adjustable
parameters: $\Delta\ped{\sigma}$ and $\Delta\ped{\pi}$, the
broadening parameters $\Gamma\ped{\sigma}$ and $\Gamma\ped{\pi}$
(that account for pair-breaking effects), the barrier transparency
coefficients $Z\ped{\sigma}$ and $Z\ped{\pi}$ (proportional to the
potential barrier height), and $w\ped{\pi}$.

Figure~\ref{Fig1}b illustrates the procedure we used to improve
this accuracy \cite{nostroPRL}. Symbols represent the experimental
normalized conductance curves of a $c$-axis In contact, measured
in zero field (``B=0'') and in the presence of a field of 1~Tesla
(``B=1T''), together with the difference between these two,
vertically shifted by 1 for best comparison (``diff.+1''). As
discussed elsewhere~\cite{nostroPRL,nostroPhysicaC}, the
$\pi$-band contribution to the experimental conductance is
strongly suppressed by the field and, at $B\simeq 1$~T, it is no
longer detectable by our fitting procedure, while the
$\sigma$-band gap remains practically unchanged. Thus, the
conductance curve measured in a field of 1~T practically contains
the $\sigma$-band conductance alone. If the conductance in zero
field is given by $\sigma\ped{B=0}=w\ped{\pi}\sigma\ped{\pi}+
(1-w\ped{\pi})\sigma\ped{\sigma}$, the curve measured in $B=1$~T
is obtained by putting $\sigma\ped{\pi}=1$ and thus has the
functional form $\sigma\ped{B=1T}=w\ped{\pi}+
(1-w\ped{\pi})\sigma\ped{\sigma}$. Consequently, the difference
between the two curves is expressed by
$\sigma\ped{diff}=\sigma\ped{B=0}-\sigma\ped{B=1T}=w\ped{\pi}(\sigma\ped{\pi}-1)$
and only contains the $\pi$-band contribution.

Fitting the experimental curves reported in the figure (symbols)
with the corresponding functions $\sigma\ped{B=0}$,
$\sigma\ped{B=1T}$ and $\sigma\ped{diff}$ gives indeed very good
results. The fitting curves are reported in Figure\ref{Fig1}b as
solid lines. The best-fitting parameters are:
$\Delta\ped{\sigma}=7.1 \pm 0.1$ meV, $\Gamma\ped{\sigma}=1.7$
meV, $Z\ped{\sigma}=0.6$ from the fit of the curve in magnetic
field; $\Delta\ped{\pi}=2.80 \pm 0.05$ meV, $\Gamma\ped{\pi}=2.0$
meV, $Z\ped{\pi}= 0.6$ from the fit of the difference. The clear
advantage of this procedure, that allows separating the partial
$\sigma$- and $\pi$-band conductances, is that the number of free
parameters in each fitting function is reduced to three
($w\ped{\pi}$ is fixed to the value one gets from the fit of
$\sigma\ped{B=0}$, i.e., in the case of Figure~\ref{Fig1}b,
$w\ped{\pi}=0.98$) with a consequent reduction of the uncertainty
on the gap amplitudes.

If this procedure is repeated at any temperature, one obtains the
complete temperature dependence of the gaps reported in
Figure~\ref{Fig1}c (symbols). Two BCS-like curves (dots) and the
$\Delta\ped{\sigma,\pi}(T)$ curves predicted by the two-band model
\cite{Brinkman} (solid lines) are reported for comparison. It is
clear that our data are so precise that they reproduce the
theoretically-predicted downward deviation of $\Delta\ped{\pi}$
from the BCS-like behaviour. However, the actual shape of the
curve does not correspond to the theoretical one, maybe indicating
that, in our samples, the values of some parameters are slightly
different from those used in the model.

\begin{figure}[t]
\begin{center}
\includegraphics[keepaspectratio, width=0.5\textwidth]{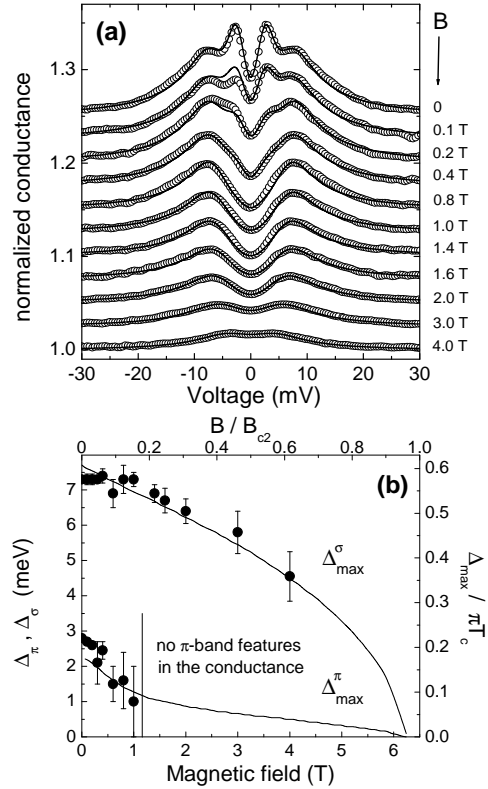}
%\vspace{-5mm}
\caption{(a) Normalized conductance curves of a Ag-paint,
$ab$-plane junction, in magnetic fields of increasing intensity
applied parallel to the $c$ axis. Symbols:experimental data;
Lines: best-fitting curves obtained with the two-band BTK model.
(b) Symbols: magnetic field dependence of the gaps as determined
from the fit of the curves in (a) (the scale is on the bottom and
left axes). Lines: dependence of the maximum pair potential,
calculated in Ref.\cite{golubov} on the reduced field
$B/B\ped{c2}$ (the scale is on the top and right axes). The
reported lines refer to the case where
$D\ped{\pi}=5D\ped{\sigma}$. }\label{Fig2}
\end{center}
%\vspace{-3mm}
\end{figure}

It is clear that the whole procedure described so far (that gave
one of the clearest direct evidences of the existence of two
distinct gaps in MgB$_2$ up to $T\ped{c}$ \cite{nostroPRL}) is
based on the disappearance of the $\pi$-band contribution to the
conductance in a field of about 1 Tesla. To obtain the whole
magnetic-field dependence of the gaps, we measured the conductance
curves of Ag-paint, $ab$-plane contacts at $T=4.2$~K, in the
presence of a magnetic field of increasing intensity applied
parallel to the $c$ axis. An example of the results is shown in
Figure~\ref{Fig2}a (symbols). The solid lines in the same graph
are the best-fitting curves of the form $\sigma(B) =
w\ped{\pi}\sigma\ped{\pi}(B) +
(1-w\ped{\pi})\sigma\ped{\sigma}(B)$, where $w\ped{\pi}$ is
assumed to be field-independent. Above 1~Tesla the best fit is
obtained by keeping $\sigma\ped{\pi}(B)=1$ (for details on the
fitting procedure see Ref.~\cite{ultimoPRL}). The values of the
gaps are shown in Figure~\ref{Fig2}b as solid symbols. For
comparison, superimposed to the experimental
$\Delta\ped{\sigma,\pi}(B)$ curves we report also the
theoretically-predicted values of the maximum order parameter
(calculated at the boundary of the vortex-lattice unit cell
\cite{golubov}) in the case where the $\pi$-band diffusivity
$D\ped{\pi}$ is 5 times greater than $D\ped{\sigma}$. The
comparison is only qualitative, since PCS rather measures an
average of the pair potential in different regions of the vortex
lattice, but the agreement is surprising. This indicates that,
even in best-quality single crystals, the $\pi$ band is in the
moderate dirty limit \footnote[7]{In this context, the expression
``clean'' and ``dirty'' refer to \emph{intraband} scattering.
According to Anderson's theorem, this scattering cannot reduce the
critical temperature of the compound nor the amplitude of the
gaps.}, in agreement with the results of STM \cite{Eskildsen} and
de Haas-van Alphen measurements \cite{Carrington}.

\begin{figure}[ht]
\begin{center}
\includegraphics[keepaspectratio, width=0.45\textwidth]{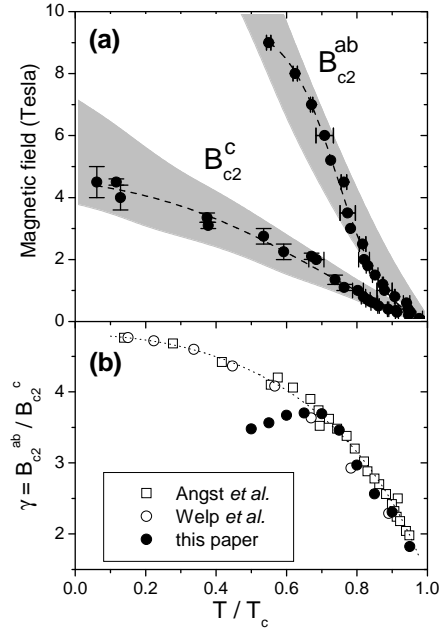}
%\vspace{-5mm}
\caption{(a) Temperature dependence of the critical field of
MgB$_2$, both for $\mathbf{B}\parallel ab$ and $\mathbf{B}
\parallel c$. Symbols: experimental data obtained from PCS; the
critical field is identified with the field that marks the return
to the normal-state conductance. The gray regions are lower
bounded by the values of $B\ped{c2}$ given by bulk-sensitive
techniques (specific heat \cite{Welp}, torque magnetometry
\cite{Angst}, thermal conductivity~\cite{Sologubenko}) and
upper-bounded by the values measured by transport (onset)
\cite{Welp,Eltsev}. (b) Temperature dependence of the anisotropy
$\gamma=B\ped{c2}^{ab}/B\ped{c2}^{c}$ measured in similar single
crystals by different techniques: torque magnetometry
($\opensquare$), specific heat ($\opencircle$), and PCS
($\fullcircle$). }\label{Fig3}
\end{center}
\vspace{-3mm}
\end{figure}

In the case of Figure~\ref{Fig2} the junction featured very good
curves, but broke down before the normal state was reached. In
many other cases we were able to follow the evolution of the
conductance curves up to the critical field \cite{ultimoPRL}, here
defined as the field that marks the return to the normal-state
conductance. Because of the anisotropy of MgB$_2$, the values of
the critical field depends very much on whether
$\mathbf{B}\parallel ab$ or $\mathbf{B} \parallel c$. The
resulting values of the critical field $B\ped{c2}$ are reported in
Figure~\ref{Fig3}a as a function of the reduced temperature
$T/T\ped{c}$.
In the same figure, the shaded regions are lower-bounded by the
values of the critical field given by bulk measurements (i.e.
thermal conductivity~\cite{Sologubenko}, specific heat~\cite{Welp}
and torque magnetometry~\cite{Angst}) and upper-bounded by the
field that, at any temperature, marks the onset of the
superconducting transition in the $\rho(B)$
curve~\cite{Welp,Eltsev}. This field has been recently
identified~\cite{Welp} with the surface critical field
$B\ped{c3}$~\cite{Hc3}. It is clearly seen that, in both the
$\mathbf{B}\parallel c$ and $\mathbf{B}\parallel ab$
configurations, our values practically coincide with the bulk
critical field at high temperatures, then approach the surface
critical field at lower temperature and finally turn again toward
the bulk values. The ``crossover'' from $B\ped{c2}$ to nearly
$B\ped{c3}$ always occurs at $T/T\ped{c}\simeq 0.7$ and thus seems
to be temperature-induced rather than field-induced. The more
pronounced curvature of our data with respect to other results
gives rise to a temperature dependence of the anisotropy ratio
$\gamma=B\ped{c2}^{ab}/B\ped{c2}^{c}$ (solid circles in
Figure~\ref{Fig3}b) which, at $T < 0.7\, T\ped{c}$, differs very
much from that given by bulk techniques (open symbols). However,
it is interesting to note that a similar $\gamma(T)$ behaviour was
theoretically predicted and reported in Ref.~\cite{Gurevich} in
the case where $D\ped{\pi}=10 D\ped{\sigma}$. Again, this
indicates that -- at least at the surface -- our MgB$_2$ single
crystals present a $\pi$ band in the moderate dirty limit, in
complete agreement with the results of Figure~\ref{Fig2}. Notice
that this conclusion has a great fundamental importance since the
role and the amount of intraband scattering in MgB$_2$ is still a
matter of debate, and alternative models (for negligible
\cite{Dahm} and non-negligible \cite{golubov,Gurevich} intraband
scattering) have been proposed to explain experimental findings in
this compound.
\begin{figure}[ht]
\begin{center}
\includegraphics[keepaspectratio, width=0.5\textwidth]{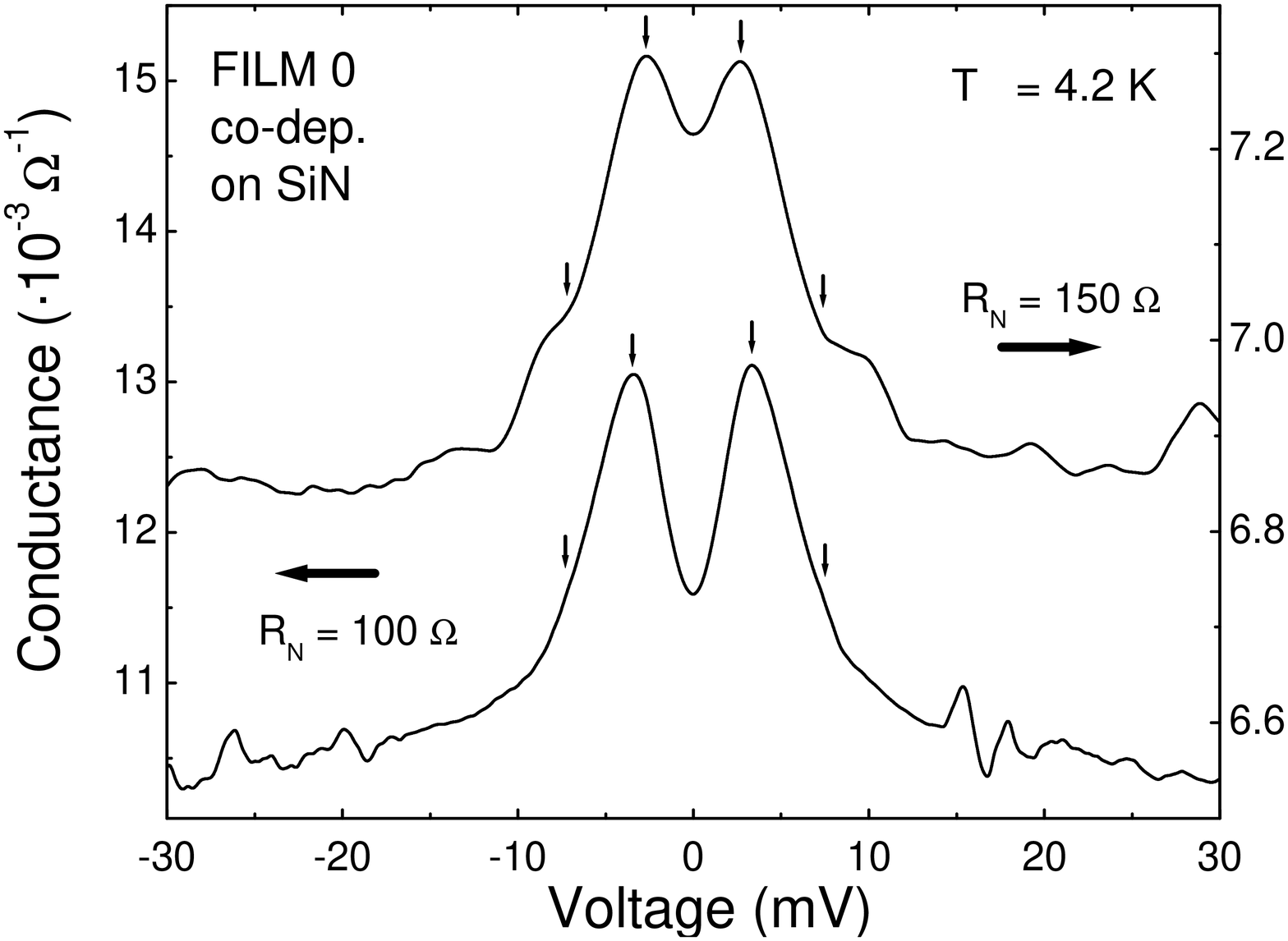}
%\vspace{-5mm}
\caption{Two examples of as-measured (i.e., non-normalized)
conductance curves of In point contacts on MgB$_2$ films grown on
SiN amorphous substrate. The curves were measured in liquid
helium. Also indicated are the positions of the conductance maxima
and of the shoulders related to the energy gaps.}\label{Fig4}
\end{center}
%\vspace{-3mm}
\end{figure}

\subsection{PCS in thin films}
Figure~\ref{Fig4} reports the conductance curves of two In/MgB$_2$
point contacts made on the surface of a thin film (``FILM 0'')
grown on silicon nitride by co-deposition of Mg and B and
subsequent \emph{in situ} annealing. The resistive critical
temperature of the film is $T\ped{c}=36$~K , and the transition is
very sharp ($\Delta T\ped{c} \leq 1$~K). The normal-state
resistance of the contacts is also indicated. The conductance
curves feature clear maxima, related to the small gap, at $\pm
2.6$~mV (upper curve) and $\pm 3.4$~mV (lower curve). A shoulder
at about $\pm 7.2$~mV is clearly seen in the upper curve, but is
also present (even if smoother) in the lower one. Comparing the
curves to those obtained in single crystals (see
Figure~\ref{Fig1}) it is clear that their shape is compatible with
a \emph{preferential} $c$-axis orientation of the films. However,
the rather evident contribution of the $\sigma$ band (especially
in the upper curve) indicates that this orientation is not
complete.

\begin{figure}[ht]
\begin{center}
\includegraphics[keepaspectratio, width=0.5\textwidth]{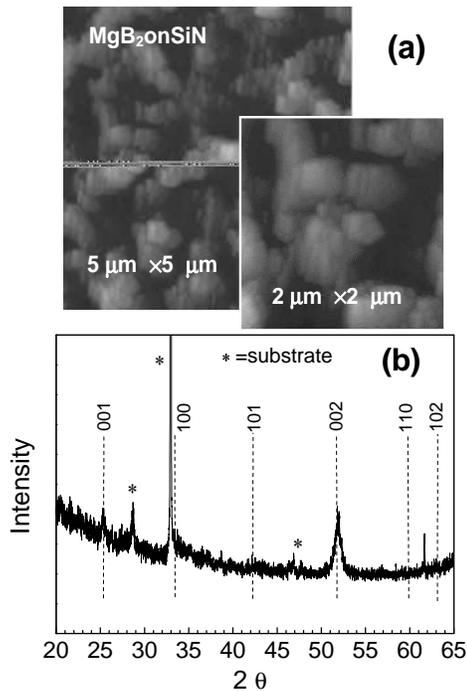}
%\vspace{-5mm}
\caption{(a) AFM images of a MgB$_2$ film grown on amorphous
silicon nitride. The presence of plate-like grains, with $c$ axis
either parallel or perpendicular to the film surface, is clearly
seen. (b) X-ray diffraction spectrum of the same film. The clear
[001] and [002] peaks, together with the absence of the other
reflections indicated, shows a preferential $c$-axis orientation.
}\label{Fig5}
\end{center}
\end{figure}
The AFM images of the film surface reported in Figure~\ref{Fig5}a
clarify this point. The main image and its enlargement (inset)
clearly show vertically- and horizontally-packed plate-like
grains, that in PCS measurements behave as a set of crystals in
which the current is injected along to the $c$ axis or along the
$ab$ planes. This explains not only the shape of the curves, with
the excess of $ab$-plane contribution, but also its variability
from point to point. An analysis of the images reported in
Figure~\ref{Fig5}a gives the average grain size ($200\div 250$~nm)
and the RMS roughness, that in this case is about 30~nm (even if
values down to 10~nm were obtained). The small roughness is
particularly important in view of the fabrication of planar
Josephson junctions, where the control of the interface smoothness
is crucial.
The XRD spectrum reported in Figure~\ref{Fig5}b shows clear (001)
and (002) diffraction maxima, coming from the $c$-axis oriented
main phase, whose intensity is however smaller than in single
crystals. The absence of a clear (101) reflection, that is the
most intense in powders, indicates that most of the film is
oriented, in spite of the amorphous SiN substrate. A very intense
spurious peak due to the crystalline Si substrate (on which SiN is
deposited) is also shown.

In addition to the information on the orientation, two other
important indications can be extracted from the Andreev-reflection
spectra reported in Figure~\ref{Fig4}.  First, the possibility to
observe Andreev reflection by PCS means that the contact is S-N,
i.e. the potential barrier $Z$ at the interface is very small
\cite{BTK}. In other words, there cannot be any insulating layer
(for example, of MgO due to the reaction with air) covering the
surface \footnote{Notice that this conclusion can be drawn here
because the contact is ``soft''. On the contrary, in the
conventional technique the tip can damage and even perforate the
surface layer and this information is lost.}. This finding is
consistent with the remarkable durability of these films that, if
kept in suitably dry atmosphere, retain their properties for
several months.
The second important indication concerns the ``clean'' or `dirty''
limit conditions at the surface of the film. In this context, the
expressions ``clean'' and ``dirty'' are referred to the
\emph{interband} scattering. According to the two-band model
appeared in literature \cite{Brinkman}, if the interband
scattering is sufficiently strong, the critical temperature falls
drastically and the two gaps merge in one single isotropic gap
$\Delta\ped{dirty}$, with BCS-like temperature dependence and
zero-temperature value $\Delta\ped{dirty}\simeq 4.2$~meV. It is
clear that the evidence of two distinct gaps in the curves of
Figure~\ref{Fig4} indicates that the surface of the films is in
clean limit, even if this kind of measurements cannot provide a
measure of the scattering rate.

\begin{figure}[ht]
\begin{center}
\includegraphics[keepaspectratio, width=0.5\textwidth]{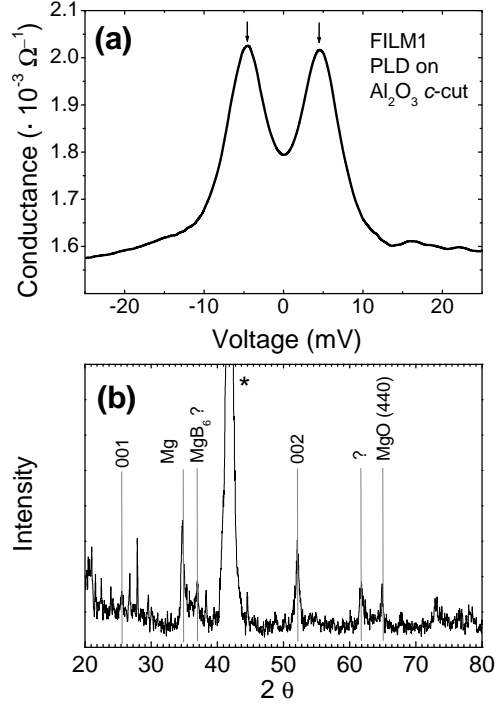}
%\vspace{-5mm}
\caption{(a) Non-normalized experimental conductance curve of a
Ag-paint contact on a MgB$_2$ thin film produced by PLD from
stoichiometric target on $c$-cut sapphire and subsequent \emph{ex
situ} annealing (``FILM 1'' in
Refs.~\cite{Ferdeghini,Ferdeghini_nuovo}). Arrows indicate the
position of the conductance maxima. (b) X-ray diffraction spectrum
of the same film, from Ref.~\cite{Ferdeghini}. Only the $(00l)$
reflections are evident for the MgB$_2$ phase, meaning that this
phase is $c$-axis oriented. Note the presence of a spurious MgO
phase, maybe related to the degradation of the surface observed by
PCS.}\label{Fig6}
\end{center}
\end{figure}

Figure~\ref{Fig6}a reports the conductance curves of a Ag-paint
spot contact on a 100~nm-thick MgB$_2$ film grown on $c$-cut
sapphire by PLD and then annealed \emph{ex situ} in Mg vapour
(``FILM~1'' in Refs.\cite{Ferdeghini,Ferdeghini_nuovo}). The
critical temperature of the film, measured by transport, is
$T\ped{c}=29.5$~K, with $\Delta T\ped{c}=2$~K. The strong $c$-axis
orientation of the film is witnessed both by the XRD spectrum
reported in Figure~\ref{Fig6}b (note that the (101) peak, that is
the most intense in powders, is absent) and by the large values of
the low-temperature anisotropy ($\gamma=3.0$), one of the highest
reported in literature for films. The film also features in-plane
orientation, with a 30$^{\circ}$-tilt with respect to the
substrate, as shown by $\Phi$-scan measurements \cite{Ferdeghini};
other characteristics are reported in Ref.\cite{Ferdeghini_nuovo}.
The conductance curve of Figure~\ref{Fig6}a shows maxima at $\pm
4.5$~mV and no clear shoulders. The absence of multigap features
and the position of the maxima suggests that the measured gap is
very likely to be $\Delta\ped{dirty}$. Since in the dirty limit
the material becomes isotropic, it is clear that, in this case,
PCS is unable to give any information about the orientation of the
film. However, it provides a direct evidence of the degradation of
the material, e.g. due to the exposition to air and moisture. This
degradation is definitely restricted to the very top layer of the
film (in fact, transport measurements gave reproducible values of
the critical field~\cite{Ferdeghini_nuovo} even if repeated after
several months) and is thus difficult or even impossible to detect
with bulk-sensitive techniques. In this sense, PCS has the unique
capability to probe the conditions of the surface, that is
particularly useful in view of the fabrication of planar Josephson
junctions, where the control of the interface is crucial.

\begin{figure}[ht]
\begin{center}
\includegraphics[keepaspectratio, width=0.5\textwidth]{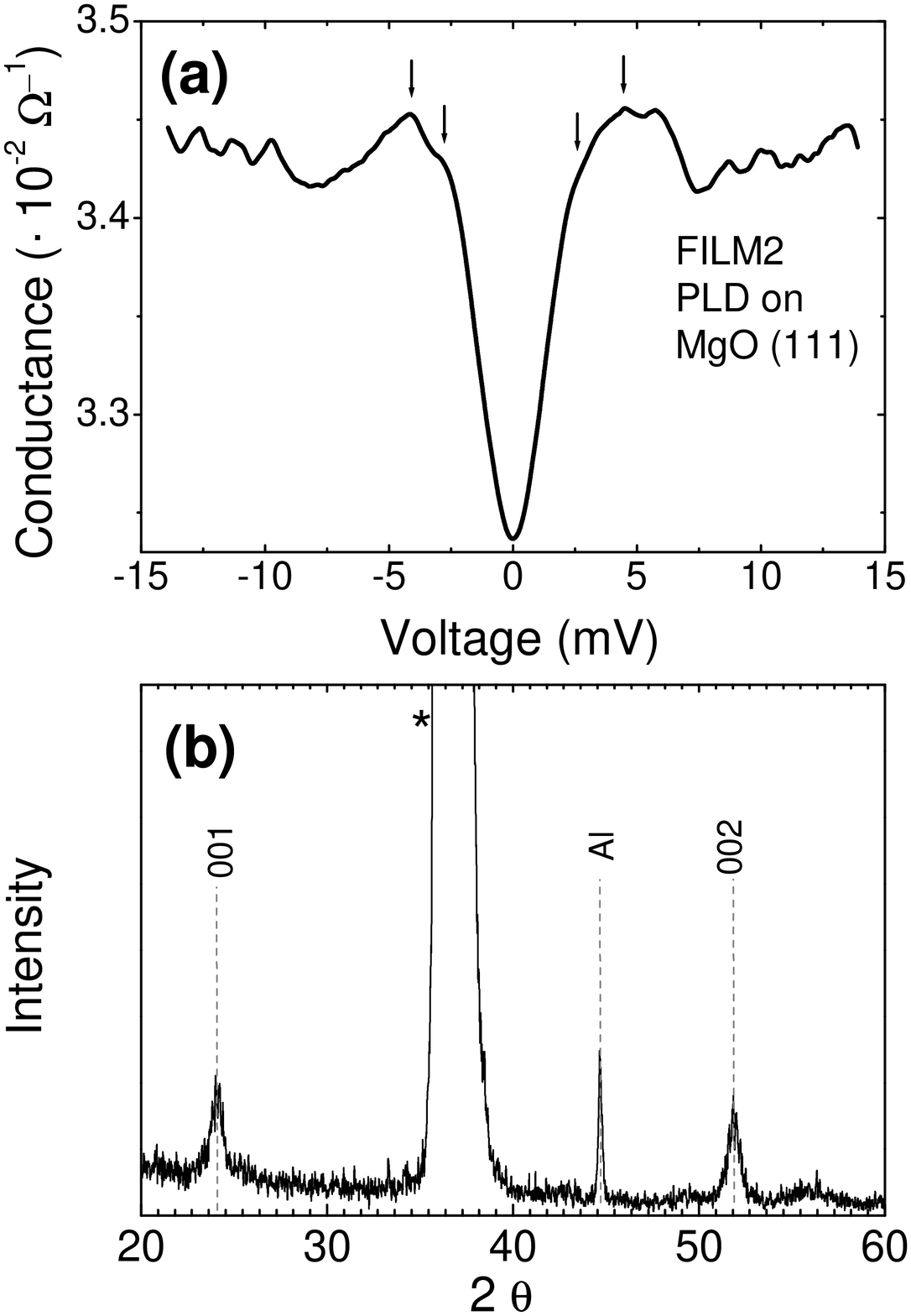}
%\vspace{-5mm}
\caption{(a) Non-normalized experimental conductance curve of a
Ag-paint contact on a MgB$_2$ thin film produced by PLD on MgO
(111) and \emph{ex situ} annealing. Arrows indicate the position
of the conductance maxima and of the shoulders related to energy
gaps. (b) X-ray diffraction spectrum of the same film, showing
clear (001) and (002) reflections. The absence of the (101) peak
was verified by tilting the sample by about 0.5$^{\circ}$ to avoid
the superposition with the substrate peak. The film is $c$-axis
oriented but features no in-plane orientation.}\label{Fig7}
\end{center}
\end{figure}

Figure~\ref{Fig7}a reports the conductance curve of a Ag-paint
contact on a MgB$_2$ film grown on MgO (111) substrate (``FILM~2''
in Ref.\cite{Ferdeghini_nuovo}). The critical temperature of the
film is a little higher than that of FILM~1, i.e. $T\ped{c}=32$~K
with $\Delta T\ped{c}=1.5$~K. As shown by the XRD spectrum
reported in Figure~\ref{Fig7}b, the film is strongly $c$-oriented,
even if it does not feature in-plane orientation. The shape of the
conductance curve clearly denotes that the contact is in the
``tunnelling'' regime (large $Z$) rather than in the
Andreev-reflection regime (small $Z$). As in the case of
Figure~\ref{Fig6}a, the curve in Fig.~\ref{Fig7}a presents
conductance maxima at energies ($\pm 4.2$~meV) that correspond to
the gap in dirty limit, $\Delta\ped{dirty}$. In this case,
however, additional features are present in the form of shoulders
at $\pm 2.7$~meV, that should correspond to the small gap
$\Delta\ped{\pi}$. This result can be explained by admitting that
parallel microcontacts are established between the Ag glue and the
film, in regions of the film where the surface is clean (and thus
$\Delta\ped{\pi}$ is observed) and dirty (where
$\Delta\ped{dirty}$ is measured). In principle, the absence of any
clear evidence of $\Delta\ped{\sigma}$ in the conductance curves
could be interpreted as a further indication of good $c$-axis
orientation of the clean material; however, in the present case
the measure is probably too noisy and broadened for this
conclusion to be definitely drawn.

As in the previous case, the observation of $\Delta\ped{dirty}$ in
some regions of the contact speaks in favour of aging effects,
that cannot be excluded due to the rather long time elapsed
between the film deposition and the PCS measurements (several
months). Further support to this hypothesis comes from the fact
that we could never observe Andreev reflection in contacts made on
this films. In other words, we always obtained S-I-N contacts,
with a non-negligible potential barrier at the interface, and with
tunnelling-like characteristics. This is probably due to the
presence of an oxide layer on the top of the film -- i.e. MgO due
to the exposition to air and moisture -- acting as an insulating
barrier and preventing the direct contact between the
superconductor and the normal metal. As a matter of fact, some
peaks due to MgO are often found in the XRD spectra, as in
Figure~\ref{Fig6} (in the case of FILM~2, however, they are not
conclusive since the substrate is made of MgO too). It is worth
noting that, again, no evidence of this surface degradation was
found in critical field measurements, that always gave a very high
value of the anisotropy ($\gamma=3.5$) supporting the
spectroscopic evidence of $c$-axis orientation.

\section{Conclusions}
In conclusion, we have shown that ``soft'' PCS, that employs small
In spots or Ag-paste drops to make the ``point'' contacts,
presents unique advantages for the investigation of the
fundamental superconducting properties of MgB$_2$ as well as for
the characterization of the morphological and electronic
properties of crystal and film surfaces. It is clear that these
advantages hold for any superconductor, provided that its surface
is sufficiently clean, stable in air, metallic and homogeneous to
obtain clear Andreev-reflection curves.  Morever, some of these
requirements are not really strict. For example, even in the
presence of an insulating surface layer (either intentionally
grown or consequent to degradation) this technique permits the
creation of useful weak-link SIN junctions (as shown in
Figure~\ref{Fig7}).

In particular, with respect to standard PCS -- where a metallic
tip is pressed against the sample surface -- our ``soft'' PCS
technique ensures: i) greater stability and reproducibility of the
junctions at the change of temperature and magnetic field; ii)
absence of damage or perforation of the surface, that prevents
unwanted contributions from the underlying bulk material; iii)
greater control of the directionality of the contact, that can be
essential in the study of anisotropic superconductors like
MgB$_2$. On the other hand, the smaller area of the contact
usually obtained by metallic tips can be useful in the study of
superconductors with non-homogeneous or granular surfaces. In this
respect, a further improvement of the technique with a reduction
of the minimum apparent contact area is under study.

This work was done within the project PRA "UMBRA" of INFM, the ASI
contract N. I/R/109/02 and the INTAS project N. 01-0617. V.A.S.
acknowledges the support from RFBR (project N. 02-02-17133) and
the Ministry of Science and Technologies of the Russian Federation
(contract N. 40.012.1.1.1357).

\end{document}